\documentclass[letterpaper, 10 pt, conference]{ieeeconf}

\IEEEoverridecommandlockouts

\usepackage{graphics} 
\usepackage{epsfig} 
\usepackage{mathptmx} 
\usepackage{times} 
\usepackage{amsmath} 
\usepackage{amssymb}  
\usepackage{booktabs}
\usepackage{subcaption}
\usepackage{makecell}
\usepackage{url}
\usepackage[symbol]{footmisc}

\title{\LARGE \bf
Quattro: \underline{T}ransformer-Accelerated Iterative Linear \underline{Qua}dratic \underline{R}egulator Framework for Fast \underline{T}rajectory \underline{O}ptimization
}

\author{Yue Wang$^{1}$, Haoyu Wang$^{1,2}$ and Zhaoxing Li$^{1}$
\thanks{*This work is supported by School of ECS, University of Southampton. The authors acknowledge the use of the IRIDIS High-Performance Computing Facility and associated support services at the University of Southampton in the completion of this work.}
\thanks{$^{1}$Yue Wang, Haoyu Wang and Zhaoxing Li are with the School of Electronic and Computer Science, University of Southampton, United Kingdom. {\tt\small \{yue.wang\}, \{haoyu.wang\}, \{zhaoxing.li\}@soton.ac.uk}.}%
\thanks{$^{2}$Haoyu Wang is also with the Department of Engineering Science, University of Oxford, United Kingdom.}%
}

\begin{document}

\maketitle
\thispagestyle{empty}
\pagestyle{empty}

\begin{abstract}
Real-time optimal control remains a fundamental challenge in robotics, especially for nonlinear systems with stringent performance requirements. As one of the representative trajectory optimization algorithms, the iterative Linear Quadratic Regulator (iLQR) faces limitations due to their inherently sequential computational nature, which restricts the efficiency and applicability of real-time control for robotic systems. While existing parallel implementations aim to overcome the above limitations, they typically demand additional computational iterations and high-performance hardware, leading to only modest practical improvements. In this paper, we introduce Quattro, a transformer-accelerated iLQR framework employing an algorithm-hardware co-design strategy to predict intermediate feedback and feedforward matrices. It facilitates effective parallel computations on resource-constrained devices without sacrificing accuracy. Experiments on cart-pole and quadrotor systems show an algorithm-level acceleration of up to 5.3$\times$ and 27$\times$ per iteration, respectively. When integrated into a Model Predictive Control (MPC) framework, Quattro achieves overall speedups of 2.8$\times$ for the cart-pole and 17.8$\times$ for the quadrotor compared to the one that applies traditional iLQR. Transformer inference is deployed on FPGA to maximize performance, achieving further up to 20.8$\times$ speedup over prevalent embedded CPUs with over 11$\times$ power reduction than GPU and low hardware resource overhead.
\end{abstract}

\section{INTRODUCTION}
Model Predictive Control (MPC) is widely used in robotics for trajectory optimization and computes optimal control inputs over future time steps \cite{bledt2018cheetah, jeon2022online, romualdi2022online, lindqvist2020nonlinear}. However, real-time performance in nonlinear settings often demands heavy computational resources \cite{nguyen2023mpc}. Differential Dynamic Programming (DDP) was reinvented in \cite{tassa2012synthesis, tassa2014control}, along with its iterative linear quadratic regulator (iLQR) variant, and is now widely used in robot control \cite{neunert2017trajectory, neunert2018whole, howell2019altro, zhu2024convergent}. Despite its strengths, iLQR still relies on a sequential pipeline that heavily demands processor clock frequency \cite{plancher2020performance}. Meanwhile, deep learning (DL) offers massive parallelism \cite{chen2020survey} and high-performance inference \cite{cai2024medusa}, which makes it well suited to address iLQR’s bottleneck. Yet, few works have combined DL with iLQR. This gap highlights an exciting path for faster, more efficient optimal control.

Although iLQR applies a Gauss-Newton approximation to linearize system dynamics, making it a more efficient but less accurate variant of DDP, it remains inherently sequential and cannot leverage multi-core computing. Multiple-shooting methods split the problem into segments \cite{pellegrini2020multiple, giftthaler2018family}, using multi-core CPUs or GPUs \cite{lee2022gpu, dai2024parallel}, but introduce extra terms to handle defects between segments \cite{li2023unified}. This typically requires more iterations to converge, reducing overall efficiency \cite{plancher2020performance}. Deep learning (DL) methods have emerged as a way to address these challenges. For example, \cite{hong2023physics} uses a neural network to approximate iLQR outputs and speed up computation, but the network does not capture key sequential dependencies and produces less reliable results. Transformers have also appeared in trajectory optimization, especially in MPC. Celestini et al. apply transformer-based models to generate near-optimal initial guesses, improving convergence and lowering computational cost \cite{celestini2024transformer}. Zinage et al. propose TransformerMPC, which predicts inactive constraints and refines initialization \cite{zinage2024transformermpc}. Although these works show promise in MPC, transformer architectures remain underexplored in iLQR. This gap presents an opportunity to better model sequential dependencies and accelerate iLQR-based control.

Deep learning has advanced rapidly over the past two decades, yielding powerful architectures for sequential data. Recurrent Neural Networks (RNNs) \cite{elman1990finding} and Long Short-Term Memory (LSTM) networks \cite{hochreiter1997long} have proven effective in many temporal tasks, but often struggle with vanishing or exploding gradients when modeling long-range dependencies \cite{lipton2015critical}. Transformers \cite{vaswani2017attention, radford2018improving, devlin2019bert} avoid these issues by using self-attention, which captures extended context without the strictly sequential updates of RNNs or LSTMs \cite{soydaner2022attention}. This design aligns well with the iterative structure of iLQR. A prime example of Transformer's potential in high-temporal tasks is Nvidia’s Deep Learning Super Sampling (DLSS), where a transformer-based approach reconstructs high-fidelity gaming frames at higher frame rates \cite{nvidia_dlss4}. This success directly motivates our pursuit of Transformer-driven iLQR solutions. Moreover, recent advances in deep learning accelerators enable efficient parallelization of Transformer architectures \cite{jouppi2017datacenter, li2020ftrans, zhao2022fpga}, paving the way for real-time, high-performance control applications.


\begin{figure*}[thpb]
    \centering
    \includegraphics[width=0.85\linewidth]{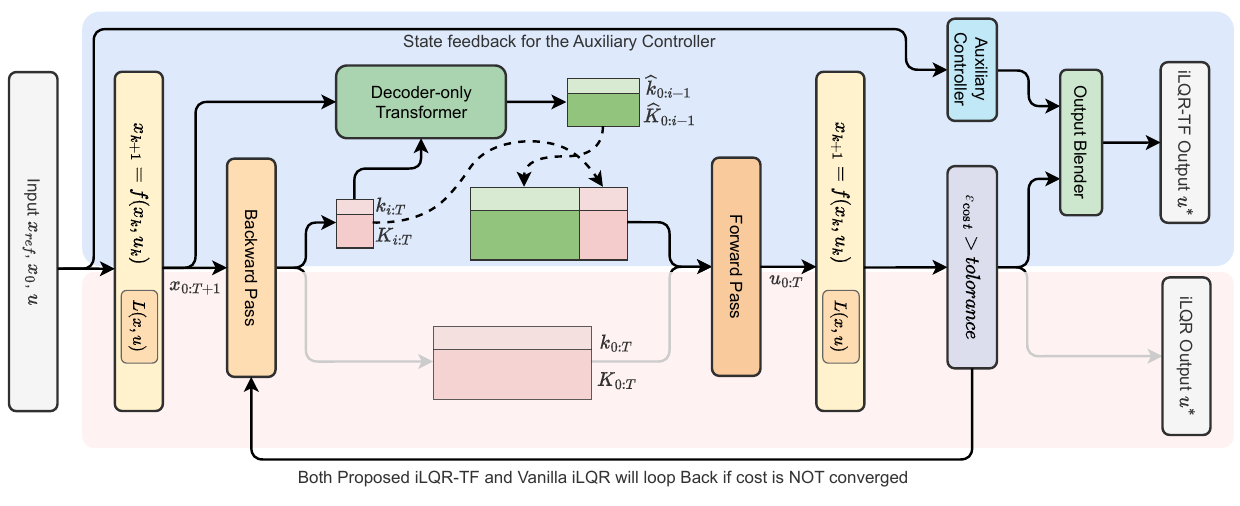}
    \caption{Comparison of standard (Vanilla) iLQR and transformer-accelerated iLQR Framework. The red-shaded area represents the standard iLQR computation process. The blue-shaded area illustrates the proposed Quattro.}
    \label{fig:compare-ilqr}
    \vspace{-0.6cm}
\end{figure*}

To address the sequential nature of iLQR computations and overcome inefficiencies in existing parallelization methods, we introduce \textbf{Quattro}\footnote[7]{Our framework is open-sourced and available at: \url{https://github.com/YueWang996/quattro-transformer-ilqr}} (Figure \ref{fig:compare-ilqr}), an iLQR framework that uses a Transformer model to generate the feedback and feedforward terms. By producing these intermediate values in parallel, Quattro reduces computation time while preserving accuracy. We validate the approach on a cart-pole \cite{tunyasuvunakool2020} and quadrotor \cite{menagerie2022github} control problems, demonstrating the effectiveness of the framework. Furthermore, an FPGA-based implementation reveals substantial performance gains, underscoring Quattro’s potential for real-time applications. The contributions of this paper are as follows:

\begin{enumerate}
\item We introduce \textbf{Quattro}, a deep acceleration framework for iLQR, intrinsically accelerated by a customized Transformer model. It significantly enhances computational efficiency through parallel computation and immediate inference compensation.
\item We validate the performance of Quattro on cart-pole and quadrotor systems on a variety of computation platforms, achieving remarkable acceleration and comparative accuracy to traditional iLQR.
\item We use the latest accelerator design framework \cite{chen2024allo} to implement the customized Transformer kernel on FPGA, achieving an optimal balance between performance, power, and hardware overhead. To our knowledge, it is the first deployment of a Transformer model on FPGA specifically for iLQR optimization.
\end{enumerate}

The remainder of this paper is organized as follows: we first introduce the background of the problem in Section~\ref{preliminary}, followed by the details of the proposed framework in Section~\ref{tfilqr}. Experiments and results are presented in Section~\ref{experiment}, and the paper concludes with Section~\ref{conclusion}.

\section{PRELIMINARIES}\label{preliminary}
\subsection{System Dynamics}
A system dynamics, or system model, describes how the state of a system changes with a given system input. The model can be defined in a discrete-time differential equation
\begin{equation}
x_{i+1} = f(x_i, u_i),
\label{eq:1}
\end{equation}
where \( x_i \) and \( u_i \) represent the state and control input at time step \( i \), respectively. A cost function can be defined to evaluate the performance of the state and input of the system over the next $N$ steps:
\begin{equation}
J(X, U) = \sum_{i=0}^{N-1} l(x_i, u_i) + l_N(x_N),
\label{eq:2}
\end{equation}
where \( l(x_i, u_i) \) is the running cost and \( l_N(x_N) \) is the terminal cost. Here, \(X:=\{x_0, x_1,\cdots,x_N\}\) denotes the state trajectory and \(U:=\{u_0, u_1,\cdots,u_{N-1}\}\) denotes the input trajectory. We can define an optimal control problem for solving the optimal \(U\) and corresponding \(X\) such that \( J(X, U) \) is minimized:
\begin{equation}
\begin{aligned}
\min_{X, U} \quad & J(X, U)\\
\textrm{s.t.} \quad &x_{i+1}=f(x_i, u_i), \quad \text{given }x_0, U_0.
\end{aligned}
\label{eq:3}
\end{equation}

\subsection{Iterative Linear Quadratic Regulator}
The iterative Linear Quadratic Regulator (iLQR) algorithm addresses nonlinear optimal control problems by repeatedly approximating the dynamics and cost around a nominal trajectory. At each iteration, the system dynamics are linearized, and the cost function is approximated up to second order. These local approximations are then used in a backward pass to compute feedback and feedforward control corrections, which subsequently refine the nominal trajectory in a forward pass. Due to the backward pass depending inherently on the results of future steps, the algorithm operates sequentially through the time horizon, inherently limiting parallel implementation. More detailed descriptions of iLQR will be presented in later sections.

\subsection{Transformer Architecture and Acceleration}

The Transformer architecture, originally developed for sequence modeling in natural language processing \cite{vaswani2017attention, dong2018speech}, relies on stacked encoder-decoder layers with self-attention mechanisms. Unlike recurrent models or sequential algorithms like iLQR, Transformers process all sequence elements simultaneously, enabling efficient parallel computation. Self-attention directly models dependencies across all time steps without recurrence, significantly shortening the information path length and making it easier to capture long-range relationships \cite{vaswani2017attention}.


This structure makes Transformers well-suited for acceleration on parallel hardware such as GPUs and Tensor Processing Units (TPUs), where matrix operations can be batched efficiently \cite{lee2022gpu}. Applying this architecture to iLQR alleviates much of the overhead from the backward pass that can limit parallelism in the standard dynamic programming approach. By learning to approximate the full control trajectory in iLQR, the transformer-based approach reduces latency and computation time, which is an advantage that is especially beneficial for real-time applications and deployment on edge devices with constrained resources.

The Transformer model is often accelerated on FPGA platforms rather than general computing units, as they are highly customizable and optimized for parallel matrix computations \cite{fuad2023survey}. As one of the Deep Learning (DL) accelerators, numerous novel architectures have been proposed and developed \cite{liu2022dynamic, park2020optimus, yang2022efa, zhao2022fpga}, alongside more agile development methodologies \cite{chen2024allo}. Notably, a new Accelerator Design Language (ADL) named Allo \cite{chen2024allo} provides a framework to directly translate Python-based transformer models into High-Level Synthesis (HLS) C code with highly optimal latency and hardware overhead. This generated HLS code can then be synthesized into Register-Transfer Level (RTL) circuits for implementation as accelerator Intellectual Property (IP) cores.

\section{TRANSFORMER-ACCELERATED ILQR} \label{tfilqr}
In this section, we revisit the iLQR computation process and explain how the Transformer enables fast and accurate iLQR computation. 
\subsection{System Rollout}
The iterative Linear Quadratic Regulator (iLQR) starts by rolling out the nonlinear system dynamics~\eqref{eq:1} from an initial state $x_0$ using an initial hypothesis of the control sequence $U_0$. This produces a nominal trajectory of states $X$. Using the obtained state-control trajectory, the performance of the trajectory can be measured through the cost function $J$ defined in~\eqref{eq:2}.

\subsection{Backward Pass}

The backward pass computes optimal feedback and feedforward control corrections. First, the nonlinear dynamics is linearized around the nominal trajectory as:
\begin{equation}
    \delta x_{i+1} \approx A_i \delta x_i + B_i \delta u_i,
    \label{eq:4}
\end{equation}
where the matrices $A_i$ and $B_i$ are Jacobians of the system dynamics with respect to the state $x_i$ and input $u_i$.

The cost is approximated to second-order around the same nominal trajectory:
\begin{equation}
    \delta l_i \approx \frac{1}{2}
    \begin{bmatrix}\delta x_i \\ \delta u_i\end{bmatrix}^\top
    \begin{bmatrix}l_{xx} & l_{xu}\\ l_{ux} & l_{uu}\end{bmatrix}
    \begin{bmatrix}\delta x_i \\ \delta u_i\end{bmatrix} 
    + 
    \begin{bmatrix}l_x \\ l_u\end{bmatrix}^\top
    \begin{bmatrix}\delta x_i \\ \delta u_i\end{bmatrix}.
    \label{eq:5}
\end{equation}

Next, following Bellman’s principle of optimality, we define a cost-to-go function $V_i(x_i)$:
\begin{equation}
    V_i(x_i) = \min_{u_i} \left[l(x_i,u_i) + V_{i+1}(f(x_i,u_i))\right],
    \label{eq:6}
\end{equation}
with terminal condition $V_N(x_N) := l_N(x_N)$.

By expanding this cost-to-go function locally up to second order, we have:
\begin{equation}
    \delta V_i(x_i) = s_i^\top \delta x_i + \frac{1}{2}\delta x_i^\top S_i \delta x_i,
    \label{eq:7}
\end{equation}
where the gradients and Hessians are defined as:
\begin{equation}
    s_i = \frac{\partial V_i}{\partial x_i}, \quad S_i = \frac{\partial^2 V_i}{\partial x_i^2}.
    \label{eq:8}
\end{equation}

We similarly expand the state-action cost $Q_i(x_i,u_i)$ as:
\begin{equation}
    \delta Q_i = \frac{1}{2}
    \begin{bmatrix}\delta x_i \\ \delta u_i\end{bmatrix}^\top
    \begin{bmatrix}Q_{xx} & Q_{xu}\\ Q_{ux} & Q_{uu}\end{bmatrix}
    \begin{bmatrix}\delta x_i \\ \delta u_i\end{bmatrix} +
    \begin{bmatrix}Q_x \\ Q_u\end{bmatrix}^\top
    \begin{bmatrix}\delta x_i \\ \delta u_i\end{bmatrix},
    \label{eq:9}
\end{equation}
where:
\begin{subequations}
\begin{align}
    Q_x &= l_x + A_i^\top s_{i+1},\\
    Q_u &= l_u + B_i^\top s_{i+1},\\
    Q_{xx} &= l_{xx} + A_i^\top S_{i+1} A_i,\\
    Q_{uu} &= l_{uu} + B_i^\top S_{i+1} B_i,\\
    Q_{ux} &= l_{ux} + B_i^\top S_{i+1} A_i = Q_{xu}^\top.
\end{align}
\label{eq:10}
\end{subequations}

Minimizing the state-action cost with respect to the input variation $\delta u_i$ yields the optimal control corrections:
\begin{equation}
    \frac{d\delta Q_i}{d u_i} = Q_u + Q_{ux}\delta x_i + Q_{uu}\delta u_i = 0,
    \label{eq:11}
\end{equation}
giving the optimal solution:
\begin{equation}
    \delta u_i^* = k_i + K_i \delta x_i,
    \label{eq:12}
\end{equation}
where:
\begin{equation}
    k_i = -Q_{uu}^{-1}Q_u, \quad K_i = -Q_{uu}^{-1}Q_{ux}.
    \label{eq:13}
\end{equation}

The backward pass computes these values recursively from the terminal state to the initial state. Additionally, the gradients and Hessians of the cost-to-go are updated recursively as:
\begin{subequations}
\begin{align}
    s_i &= Q_x + K_i^\top Q_{uu} k_i + K_i^\top Q_u + Q_{ux}^\top k_i,\\
    S_i &= Q_{xx} + K_i^\top Q_{uu} K_i + K_i^\top Q_{ux} + Q_{ux}^\top K_i.
\end{align}
\label{eq:14}
\end{subequations}
\vspace{-0.2cm}
\subsection{Forward Pass}
In the forward pass, the control sequence is updated using a line search with step size $\alpha$:
\begin{equation}
    u_i^{new} = u_i + \alpha k_i + K_i (x_i^{new} - x_i),
    \label{eq:15}
\end{equation}
where $x_i^{new}$ is the newly rolled-out state during this forward pass. This produces a new trajectory and updated cost $J^{new}$. 

After the forward pass, convergence is checked by comparing the change in the cost function. If the improvement is smaller than a pre-specified threshold (e.g., $|J^{new}-J| < \epsilon$), the iteration stops. Otherwise, the procedure repeats the backward and forward passes until convergence or a preset maximum iteration limit is reached. In practical implementations, several step sizes may be tested in parallel: each candidate produces a new control sequence and a forward pass is carried out for each, thereby preventing repeated forward-pass computations if a particular step size proves unsuccessful \cite{lee2022gpu}.

\begin{figure}[thpb]
    \centering    \includegraphics[width=0.95\linewidth]{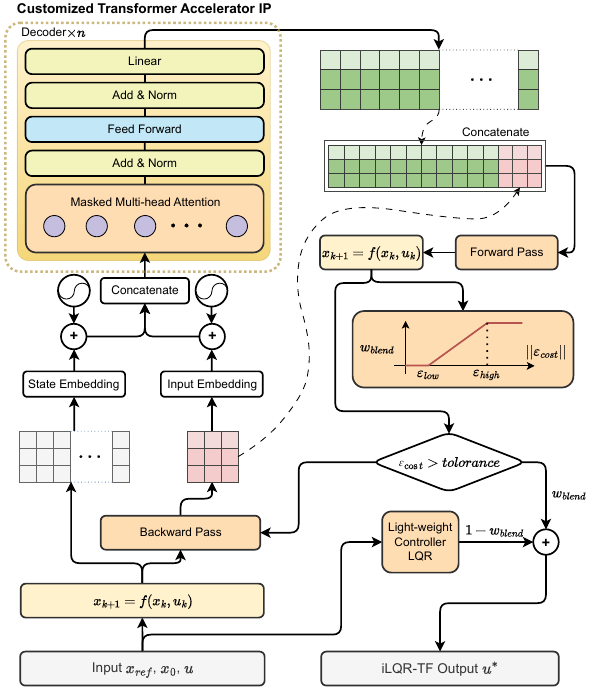}
    \caption{Detailed framework architecture of the iLQR-TF.}
    \label{fig:second-pdf}
    \vspace{-0.5cm}
\end{figure}

\subsection{Integration of Transformer} \label{sec:tf-embed}
The backward pass recursively computes the gains and updates the system input. The core idea to accelerate the algorithm is to reduce the sequential computation. In iLQR, the recursive backward pass is most time-consuming due to the extensive derivative of system dynamics and cost functions \cite{plancher2020performance}.

One idea is to replace the entire backward pass with a Transformer module \(\pi_\theta\) as in
\begin{equation}
    \hat{k}, \hat{K} = \pi_\theta (X).
\end{equation}
While it can significantly accelerate the computation because the sequential computation is fully converted to parallel computing, this purely end-to-end approach may lack robustness since it entirely omits crucial second-order cost-to-go and sensitivity information from the backward pass. Incorporating explicit backward pass information (e.g., cost-to-go Hessians or gradients) as inputs to the Transformer can enable more informed and reliable predictions. 

Therefore, in our framework (as the Figure \ref{fig:second-pdf} shown), instead of computing the full backward pass, we perform a partial backward pass which computes the gain matrices $k_{i: T-1}=\{k_i, \cdots, k_{T-1}\}$ and $K_{i: T-1}=\{K_i, \cdots, K_{T-1}\}$. The rest of the gain matrices are predicted by the Transformer model:
\begin{equation}
    \hat{k}_{0:i-1}, \hat{K}_{0, i-1} = \pi_\theta (X, k_{i: T-1}, K_{i: T-1}).
\end{equation}
By doing this, it preserves the essential structure of optimal control thus improving the robustness and stability of the overall solution.

As in Figure \ref{fig:second-pdf}, a decoder-only Transformer is adopted for predicting iLQR gain matrices due to its causal structure, which naturally aligns with the sequential and temporal characteristics inherent to gain computations in optimal control. An encoder-based Transformer is less suitable because its bidirectional attention structure does not preserve temporal causality, which is crucial for sequential control tasks. Similarly, a full encoder-decoder Transformer introduces unnecessary complexity and computational overhead without clear advantages, as the primary task here is modeling sequential dependencies rather than mapping between distinct input-output sequences. Thus, the decoder-only Transformer provides a structurally more appropriate choice for modeling the causal and sequential nature of the iLQR gain prediction.

In order to feed both state and gain matrices, we add an additional channel to the Transformer module. The gain matrices are stacked as a single input matrix. After the state embedding and gain embedding, they are positionally encoded and concatenated before being fed to the multi-head attention. The output of the Transformer is a flattened vector which is reshaped to extract the gain matrices $\hat{k}_{0:i-1}$ and $\hat{K}_{0, i-1}$. After concatenating the predicted gains and the calculated gains, the full feedback and feedforward gains are obtained and used to compute the forward pass. 

\vspace{-0.2cm}
\subsection{Optional Output Blender}

To avoid potential oscillations caused by small residual gains from the transformer-based controller (iLQR-TF) near equilibrium, we incorporate an \textit{optional} lightweight Linear Quadratic Regulator (LQR) around the reference state. A standard blending mechanism smoothly transitions between the iLQR-TF and LQR outputs based on the current cost \( J \). Defining two thresholds (\(\varepsilon_{\text{low}}\), \(\varepsilon_{\text{high}}\)), the blending weight is computed as:
\begin{equation}
w_{\text{blend}} = 
\left\{
\begin{aligned}
&0, && \|J\| \leq \varepsilon_{\text{low}},\\[3pt]
&\frac{\|J\| - \varepsilon_{\text{low}}}{\varepsilon_{\text{high}} - \varepsilon_{\text{low}}}, && \varepsilon_{\text{low}} < \|J\| < \varepsilon_{\text{high}},\\[3pt]
&1, && \|J\| \geq \varepsilon_{\text{high}}.
\end{aligned}
\right.
\end{equation}

The final blended control output is then:
\begin{equation}
u^* = w_{\text{blend}}\,u_{\text{iLQR-TF}} + (1 - w_{\text{blend}})\,u_{\text{LQR}},
\end{equation}
which implies that the system is fully controlled by iLQR-TF when \( w_{\text{blend}} = 1 \), and entirely by LQR when \( w_{\text{blend}} = 0 \).

\section{EXPERIMENTS AND ANALYSIS} \label{experiment}

We evaluated our iLQR-Transformer framework on two benchmark control problems: a cart-pole system and a quadrotor, both simulated using MuJoCo \cite{todorov2012mujoco}.

\subsection{Data Collection}
We generated training data by solving optimal control problems using iLQR. Specifically, each system was initialized from diverse initial states, and standard iLQR was performed to compute the gains \((k, K)\) and input sequence \(U\). We incorporated iLQR within an MPC framework, collecting \((X, k, K, U)\) for each iLQR iteration at every MPC step.

For the low-dimensional cart-pole system \(\left(x \in \mathbb{R}^4\right)\), we discretized the dynamics at a 0.01 s time step and simulated for 15 s. Initial x-positions and angles were sampled on a grid over \([-0.5, 0.5]\) in 0.05 increments.

In contrast, for the higher-dimensional quadrotor system \(\left(x \in \mathbb{R}^{12}\right)\), grid search is impractical. Instead, we employed Latin Hypercube Sampling (LHS) \cite{mckay2000comparison} to draw 2,000 initial states. These spanned \(\text{pos}_x, \text{pos}_y \in [-0.3, 0.3]\), \(\text{pos}_z \in [0.2, 0.5]\), \(\text{roll}, \text{pitch} \in [-0.2, 0.2]\), and \(\text{yaw} \in [-0.5, 0.5]\), with all velocities initially set to zero. This ensures broad coverage of the higher-dimensional state space while restricting the initial velocities. We used the same simulation approach to collect iLQR computation results. 

\subsection{Transformer Model and Accelerator Design}
Transformer models were implemented using PyTorch and trained using the high-performance computing cluster. To address the differing complexities between the two control systems, model parameters were adjusted accordingly. In the case of the cart-pole system, a three-layer decoder-only architecture was employed, with each layer incorporating 4 attention heads and a model dimensionality of 128. The dimension of the feedforward layer is 256, which captures the features of the input sequence. For the quadrotor system, we choose the same Transformer architecture and parameters. However, we extend the feedforward layer from 256 to 512, to capture more input features and dependencies due to the complicated dynamics. 


We primarily adopted the recent agile Transformer ADL framework proposed in \cite{chen2024allo} to build an efficient and practical hardware design. This approach brings together several essential kernel components that form the full Transformer computation. By taking advantage of the parallel matrix multiplication capability of FPGAs, we focused the hardware acceleration on the main computation blocks, including \textbf{multi-head attention, layer normalization, residual connections, and the feed-forward network.} These are the most computationally intensive parts of the decoder. Other operations, such as embedding, positional encoding, and matrix combination, were handled by the CPU since they offer limited opportunities for parallel processing. To balance latency and hardware resource usage, we applied several HLS directives such as pipeline, unroll, partition, and the use of dual-port RAM during the design process.

\begin{table}[htbp]
\caption{FPGA Development Experiment Setup}
\label{tab:fpga_setup}
\centering
\begin{tabular}{ll}
\toprule
\textbf{Item} & \textbf{Specification} \\
\midrule
FPGA Device Series & AMD Kintex Series FPGA \\
Accelerator Design Language & Allo~\cite{chen2024allo} \\
Main Programming Languages & Python, HLS C, Verilog HDL \\
EDA Tools & Vitis, Vivado \\
Clock Frequency & \makecell[l]{80 MHz (Cart-Pole)\\200 MHz (Quadrotor)} \\
\bottomrule
\end{tabular}
\vspace{-0.3cm}
\end{table}


\subsection{Prediction Accuracy} \label{sec:accuracy}
\begin{figure}[thpb]
\vspace{-0.2cm}
    \centering
    \includegraphics[width=0.85\linewidth]{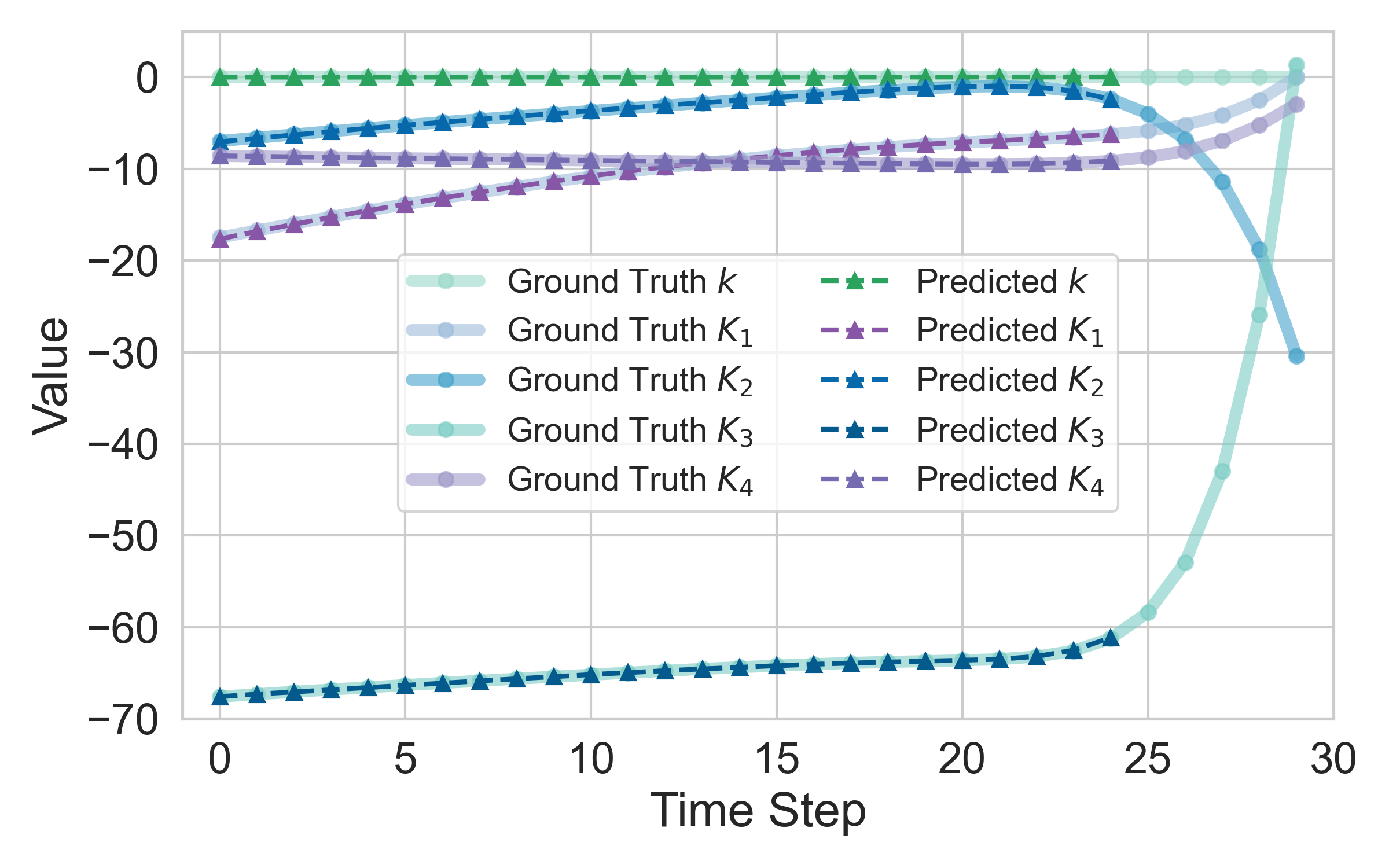}
    \caption{Comparison between ground truth (solid lines) and transformer-predicted (dashed lines) gain sequences for the cart-pole system, using a prompt length of 5 prior gain steps.}
    \label{fig:error-compare}
    \vspace{-0.2cm}
\end{figure}

For the cart-pole system, we regulate the cart to stop at position 0 along the x-axis and stabilize the rod angle at 0 radians. Figure \ref{fig:error-compare} shows the predicted gain matrices $\hat{k}$ and $\hat{K}=\{K_1, K_2, K_3, K_4\}$ closely matching the actual gains computed from iLQR. Each element $K_i$ corresponds to a feedback gain matrix $K$ component at a given time step. 

\begin{figure}[thpb]
    \centering
    \begin{subfigure}[b]{\linewidth}
        \centering
        \includegraphics[width=\linewidth]{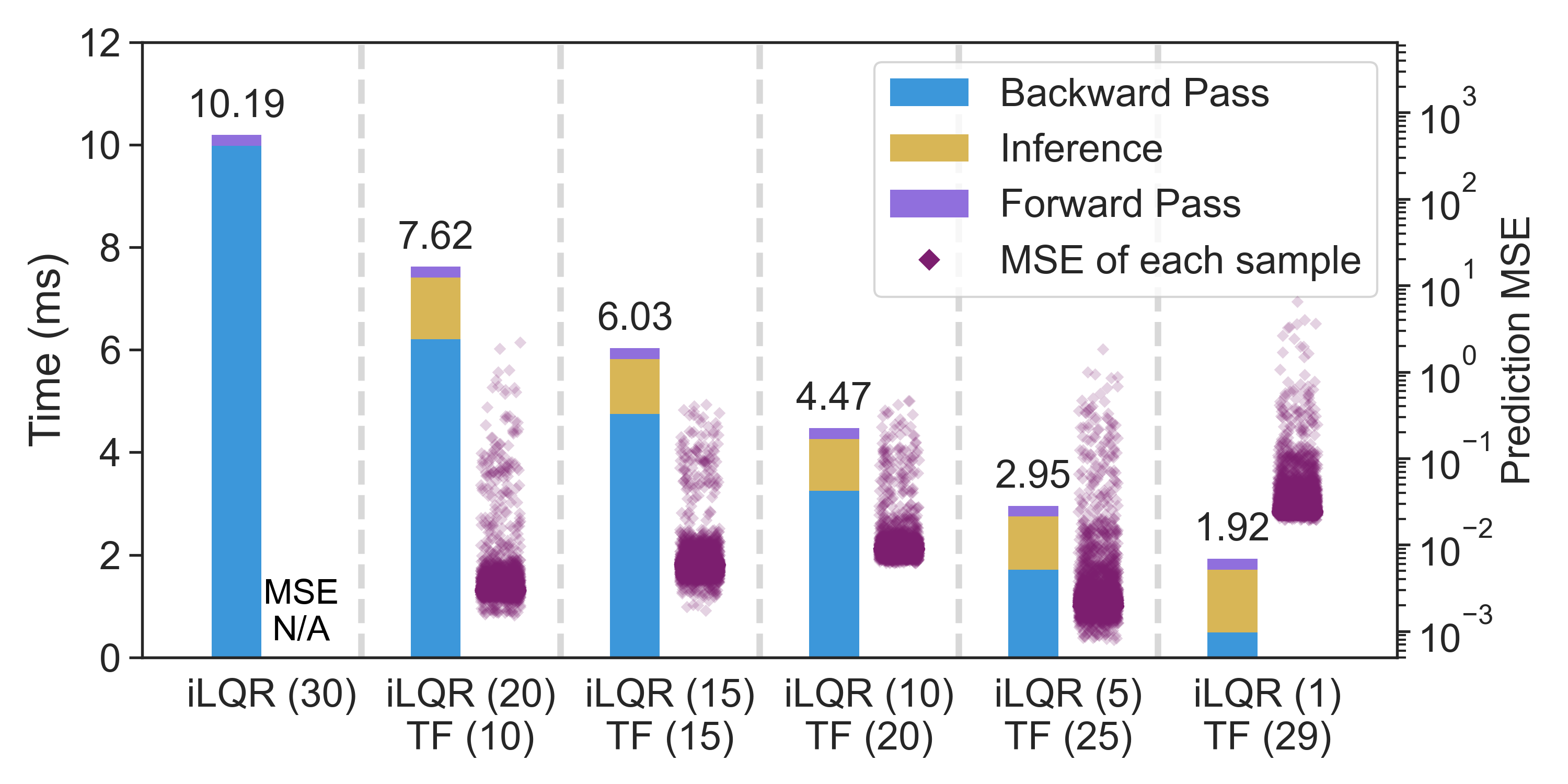}
    \end{subfigure}
    
    \begin{subfigure}[b]{\linewidth}
        \centering
        \includegraphics[width=\linewidth]{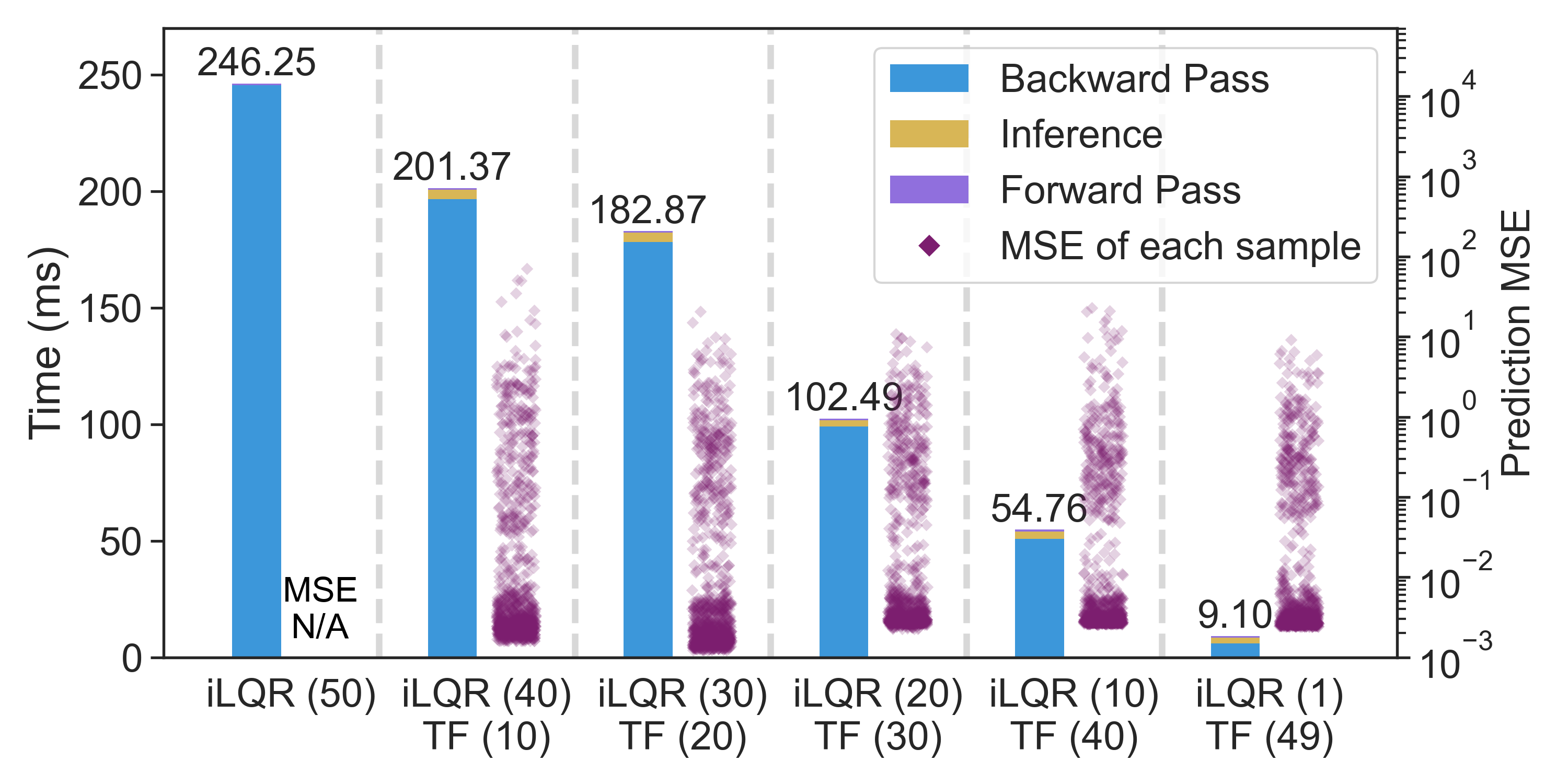}
        \label{fig:time-mse-quadrotor}
    \end{subfigure}
    \vspace{-1cm}
    \caption{Running time and prediction accuracy for different computational distributions between iLQR and Transformer for the cart-pole (upper) and the quadrotor (lower) systems. Bar plots show computation time per iteration (left Y axis), and scatter plots indicate MSE distributions (right Y axis) for varying prompt lengths. Numbers in parentheses (e.g., iLQR (30) TF (20)) denote the computational steps performed by iLQR and the Transformer, respectively.}
    \label{fig:time-mse-comparison}
    \vspace{-0.4cm}
\end{figure}

We further evaluated prediction accuracy with different prediction lengths of the intermediate gain matrices. For a specific horizon, iLQR performed a partially backward pass as Section \ref{sec:tf-embed} described, and the transformer model predicted the rest of the gain sequence to obtain the updated input sequence $U_{iLQR-TF}$ by a forward pass. We measured the MSE between this $U_{iLQR-TF}$ and the full iLQR computational result. The dotted plot in Figure \ref{fig:time-mse-comparison} presents MSE for different payload allocations between iLQR and Transformer. For the cart-pole system ($T=30$), the MSE values for each sample are relatively centralized and gradually increase as the number of computed iLQR gain matrices decreases.

For the quadrotor system, with a prediction horizon $T=50$, the Transformer inputs consist of state sequences $X\in \mathbb{R}^{T\times 12}$ and stacked gain matrices $K_{stacked} \in \mathbb{R}^{T\times 52}$ (reshaped from $k\in \mathbb{R}^{T\times 4}$ and $K\in \mathbb{R}^{T\times 4\times12}$). Figure \ref{fig:time-mse-comparison} (lower plot) illustrates MSE, showing higher and more dispersed values due to the increased complexity and dimensionality compared to the cart-pole system. However, despite the higher dispersion resulting from the squared error amplification across larger gain matrices, most prediction errors remain very low (between $10^{-1}$ and $10^{-3}$). Additionally, for varying prediction lengths, the distribution of MSE samples remains consistent, indicating that a higher proportion of transformer-predicted data can be utilized to maximize parallelization and improve overall computational efficiency.

\vspace{-0.2cm}
\subsection{Computation Speedup and Accelerator Performance}
We first tested our framework on an Apple Silicon M4 Pro (10-core CPU). Figure \ref{fig:time-mse-comparison} illustrates computation times for varying allocations between iLQR and the Transformer. As we reduce the computational steps executed by iLQR, overall execution time notably decreases, leveraging the highly parallel computation of Transformer. For the cart-pole system, average computation per iteration reduces from 10.19 ms to 1.92 ms \textbf{(5.3$\times$ faster)}. For the more complex quadrotor, iteration time decreases significantly from 246.25 ms to 9.10 ms \textbf{(27$\times$ faster)}, highlighting our method's advantage in handling larger, computationally intensive systems.

\begin{figure}[thpb]
    \centering
    \begin{subfigure}[b]{\linewidth}
        \centering
        \includegraphics[width=\linewidth]{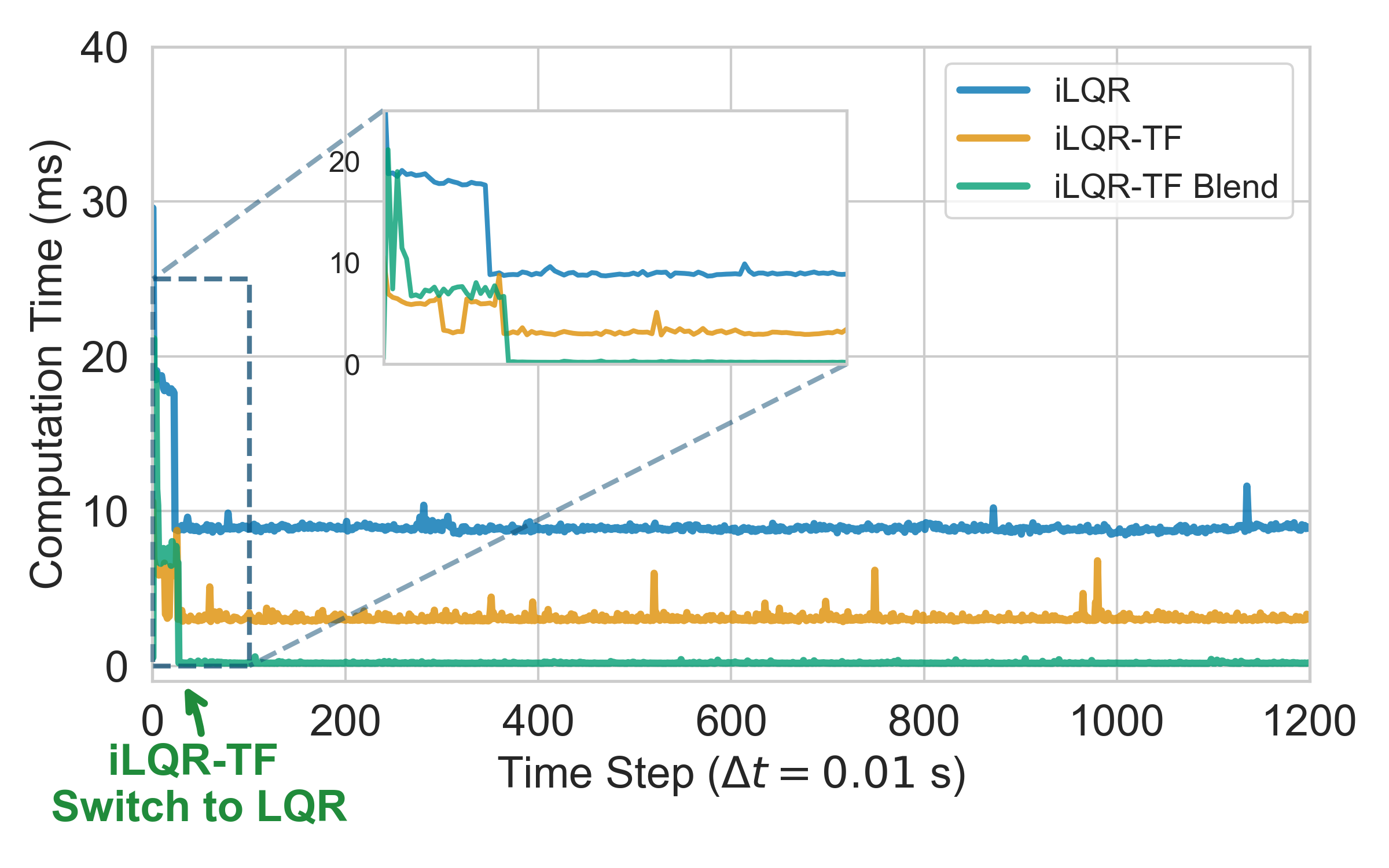}
    \end{subfigure}

    \begin{subfigure}[b]{\linewidth}
        \centering
        \includegraphics[width=\linewidth]{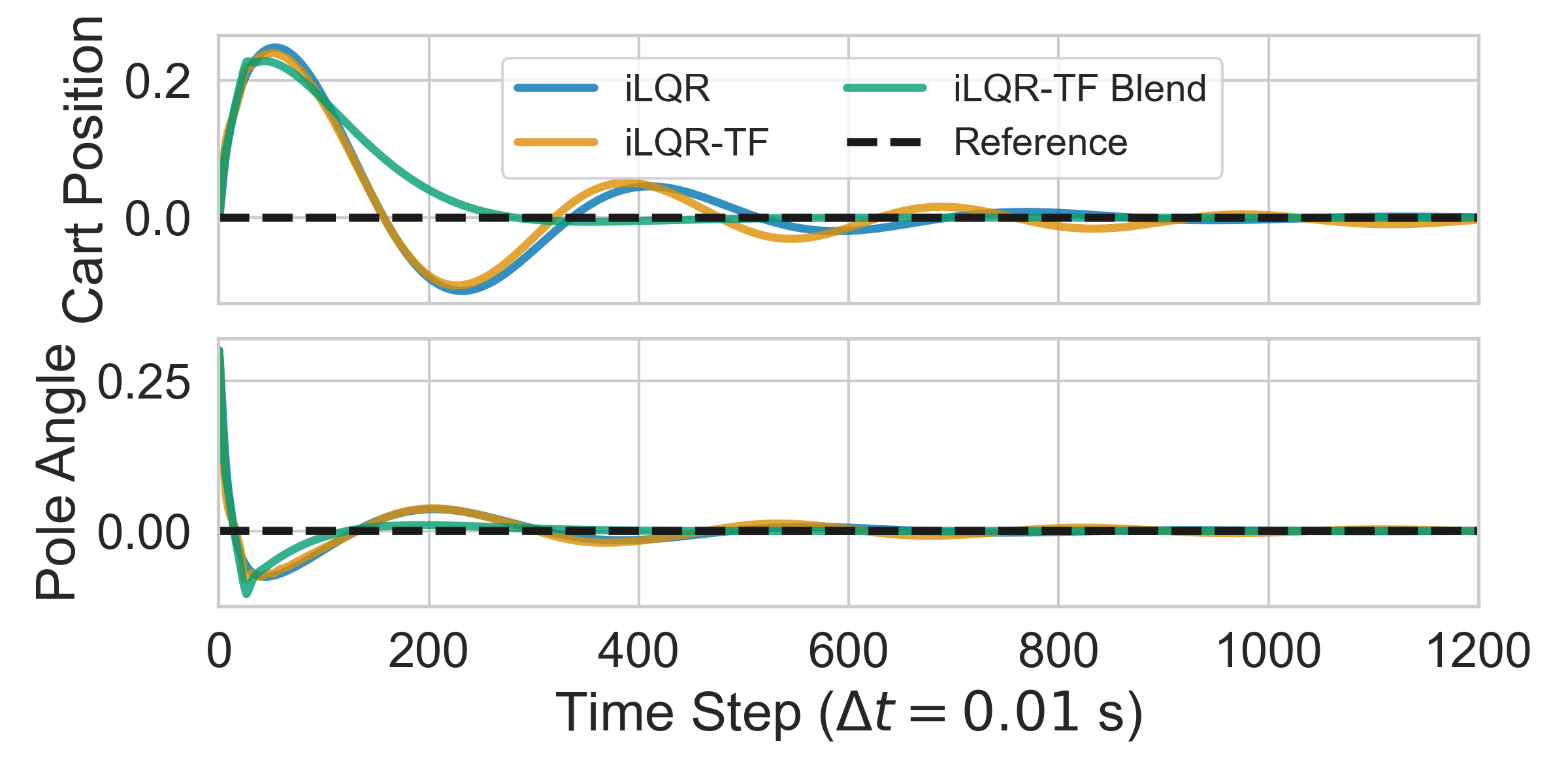}
    \end{subfigure}
    \vspace{-0.5cm}
    \caption{Cart-pole system: (upper) computational efficiency and (lower) system state trajectories comparison among iLQR, iLQR-TF, and blended (mixture of iLQR-TF and LQR). The horizon is 30 and the Transformer prediction length is 25. }
    \label{fig:cartpole-computation}
    \vspace{-0.4cm}
\end{figure}

Compared to another accelerated iLQR algorithm in \cite{plancher2020performance}, which typically requires additional iterations to converge, our method achieves a comparable number of iterations as vanilla iLQR to reach the optimal solution, thereby demonstrating significant overall time savings. As depicted in the upper plot of Figure \ref{fig:cartpole-computation}, the total simulation time for the cart-pole system is reduced from 10.5 s to 3.7 s when using 5 iLQR steps and 25 Transformer-assisted steps, resulting in a \textbf{2.8$\times$ speedup}. This aligns closely with the observed time savings illustrated in Figure \ref{fig:time-mse-comparison} (upper). In the quadrotor control scenario, for a simulation with 10,000 total simulation steps and 500 MPC control computations, our method substantially reduces the computation time from 237 s (pure iLQR computation at every MPC step) to 13.3 s by combining 1 step of iLQR computation with 49 subsequent steps predicted by the Transformer. This leads to a \textbf{17.8× speedup}. Additionally, when compared to state-of-the-art OCP solvers OSQP, ECOS, and SCS reported in \cite{cvxpy}, our approach achieves speedups of 2.49×, 3.63×, and 2.4×, respectively.

Table \ref{tab:fpga_performance} shows that our accelerator delivers the expected performance in Transformer inference. In the cart-pole control scenario, it achieves similar speed to the Apple M4 Pro@3.5 GHz, while running \textbf{1.55$\times$ faster than the NVIDIA RTX4070 and 17.67$\times$ faster than the Cortex-A72 in the Raspberry Pi}. In the quadrotor control scenario, our accelerator is \textbf{1.8$\times$ faster than the M4 Pro CPU and 20.8$\times$ faster than the Raspberry Pi}. Although the inference time is close to that of the RTX4070 in the quadrotor task, our accelerator uses much less power, consuming only 1.15 W for the cart-pole and 1.68 W for the quadrotor, compared to the steady 13 W drawn by the GPU in both cases.


The hardware resource overhead is extremely low on a mid-range FPGA, with average utilization remaining below 10\% for all resource types except BRAM, which is primarily used to store model parameters. Under the same 20nm process, the estimated silicon area is significantly smaller than that of typical CPUs or GPUs.
\vspace{-0.2cm}
\begin{table}[htbp]
\caption{FPGA Accelerator Performance Metrics}
\label{tab:fpga_performance}
\centering
\begin{tabular}{lcc}
\toprule
\textbf{Metric} & \textbf{Cart-Pole} & \textbf{Quadrotor} \\
\midrule
\multicolumn{3}{l}{\textit{Inference Performance}} \\
Latency (ms) & 1.05 & 1.73 \\
Speed-Up (vs. 10-core M4 Pro CPU) & 0.99$\times$ & 1.8$\times$ \\
Speed-Up (vs. RTX4070 GPU) & 1.55$\times$ & 1.01$\times$\\
Speed-Up (vs. Quad-core Cortex-A72) &17.67$\times$ & \textbf{20.8$\times$} \\
Power Reduction (vs. RTX4070 GPU) &  \textbf{11.3$\times$} & 7.74$\times$ \\
\addlinespace
\multicolumn{3}{l}{\textit{Resource Utilization}} \\
BRAM Usage (\%) & 24 & 41 \\
DSP Usage (\%) & 4 & 9 \\
FF Usage (\%) & 4 & 3 \\
LUT Usage (\%) & 9 & 8 \\
\bottomrule
\end{tabular}
\end{table}
\vspace{-0.3cm}
\subsection{Trajectory Tracking Performance}
Our framework effectively controls both systems to follow desired trajectories. Figure \ref{fig:cartpole-computation} (lower plot) demonstrates that the transformer-assisted (iLQR-TF) controller closely matches the ground truth iLQR controller's performance in stabilizing the cart-pole system. Introducing LQR as the system approaches a linear region enables smooth convergence to the desired state.

\begin{figure}[thpb]
\centering
\includegraphics[width=0.9\linewidth]{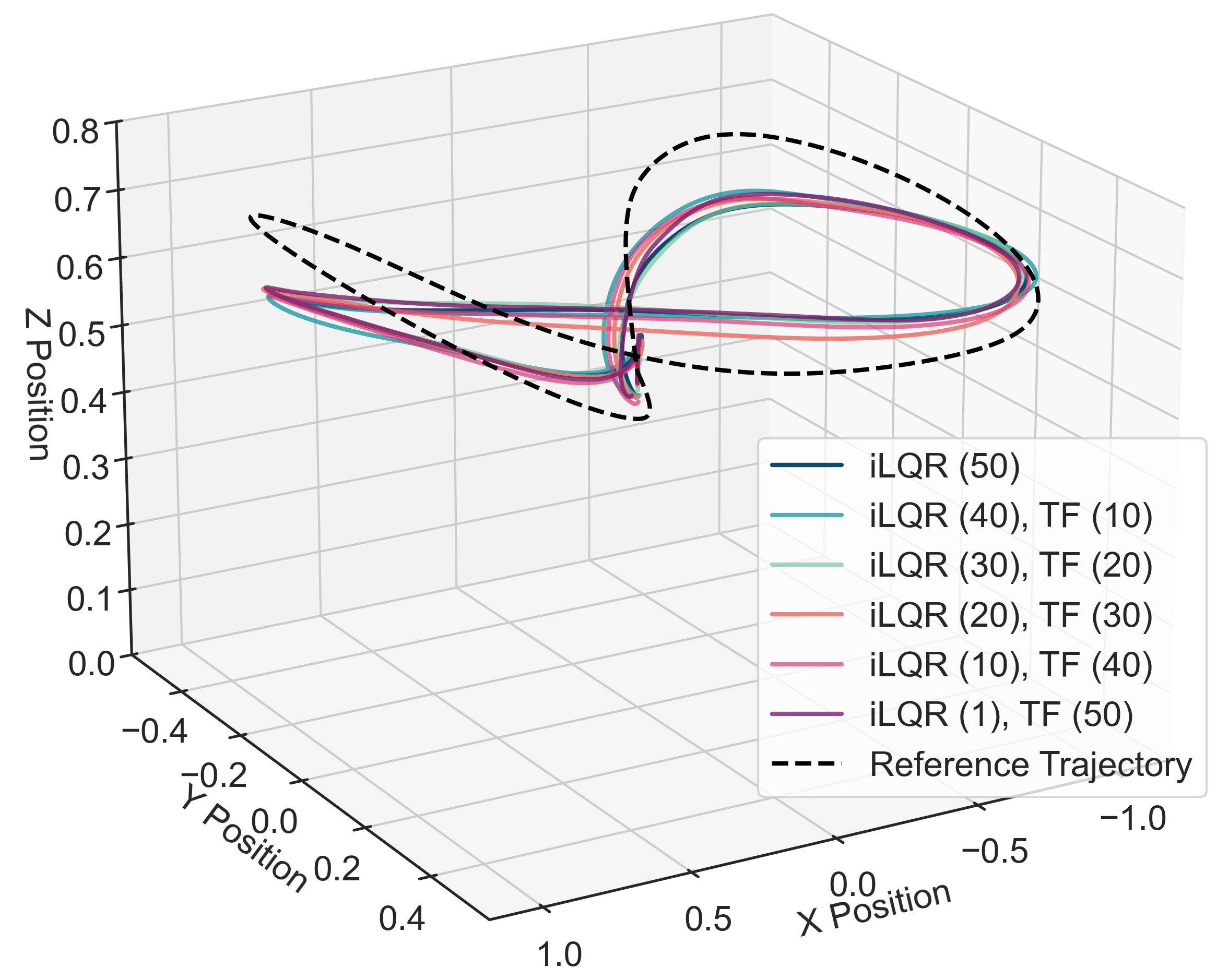}
\caption{Quadrotor trajectory tracking performance under varying allocations of iLQR and Transformer computation steps, demonstrating accurate tracking relative to the reference trajectory.}
\label{fig:quad-traj}
\vspace{-0.4cm}
\end{figure}

We also tested trajectory tracking with a more complex 8-shaped trajectory for the quadrotor system using our iLQR-TF approach. As shown in Figure \ref{fig:quad-traj}, the transformer-assisted control accurately tracks the reference trajectory across various prediction lengths. The predicted trajectories consistently align closely with the ground truth computed by full-horizon iLQR (50 steps).

\subsection{Hyperparameter Exploration}

The accuracy of the Transformer model predictions depends on the lengths of the input sequences for both states and gains. As shown in Figure~\ref{fig:time-mse-comparison}, performance varies with the number of initial iLQR steps before switching to the Transformer. To balance prediction accuracy and inference time, we chose 5 iLQR steps for the cart-pole system and 1 step for the quadrotor.

\begin{figure}[thpb]
    \centering
    \begin{subfigure}[b]{\linewidth}
        \centering
    \includegraphics[width=\linewidth]{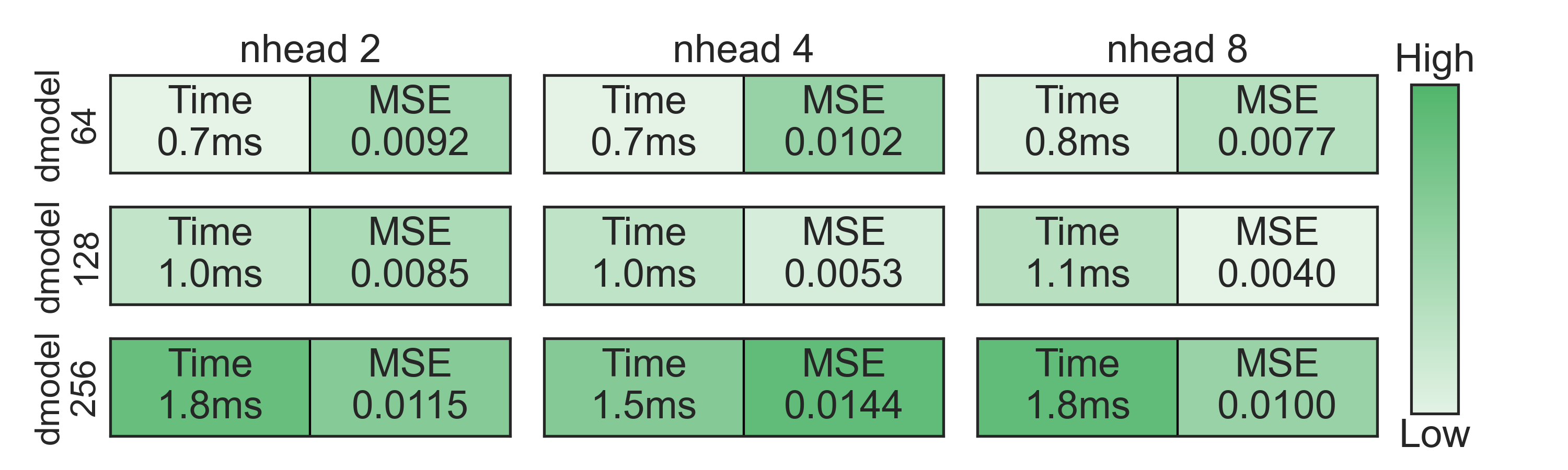}
    \end{subfigure}
    
    \begin{subfigure}[b]{\linewidth}
        \centering
        \includegraphics[width=\linewidth]{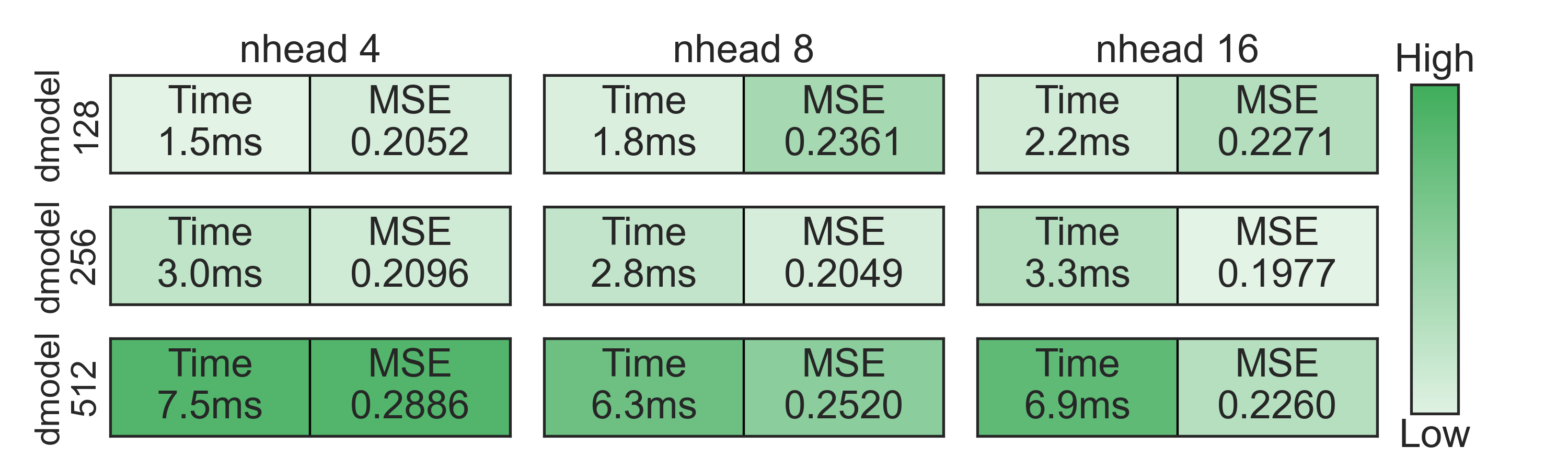}
    \end{subfigure}

    \caption{Hyperparameter exploration of Transformer models on cart-pole (upper) and quadrotor (lower). Each cell shows MSE and inference time for different \texttt{dmodel} and \texttt{nhead} settings. The lighter green cell indicates better overall performance.}
    \label{fig:ablation}
\end{figure}

We further explored key Transformer hyperparameters, including the number of attention heads (\texttt{nhead}) and the model dimension (\texttt{dmodel}), to validate our design choices. Figure~\ref{fig:ablation} presents the results. For the cart-pole system, the configuration \texttt{dmodel}=128 and \texttt{nhead}=4 achieved the lowest MSE (0.0053) with a fast inference time (0.9ms). For the quadrotor, the same configuration yielded the lowest inference time (2.0ms) and a good MSE (0.2051), highlighting a consistent trade-off across tasks.

As shown in Figure~\ref{fig:ablation}, smaller models generally perform better in both accuracy and speed, while increasing \texttt{dmodel} or \texttt{nhead} beyond a certain point leads to diminishing returns or higher latency. The selected configuration balances model capacity with runtime efficiency: \texttt{dmodel}=128 provides sufficient expressiveness without excessive cost, and \texttt{nhead}=4 allows effective multiple attention. These results confirm the suitability of lightweight architectures for real-time deployment.

\vspace{-0.8cm}
\section{CONCLUSION} \label{conclusion}

In this paper, we presented Quattro, a Transformer-accelerated iLQR framework that addresses the sequential bottleneck in traditional iLQR algorithms through parallel inference of control gains. By predicting intermediate computational variables, Quattro maintains the structure of optimal control while significantly reducing computation time. Our experiments on cart-pole and quadrotor systems demonstrate that Quattro achieves up to 27$\times$ per-iteration speedup and 17.8$\times$ end-to-end acceleration within an MPC framework. FPGA-based deployment further shows 17–20$\times$ speedups over edge CPUs and up to 1.55$\times$ than GPU with 11.3$\times$ power reduction and dramatically low hardware overhead. While larger Transformer models can yield even higher prediction accuracy (sometimes surpassing the original iLQR), they introduce higher latency and hardware overhead. Therefore, we select a lightweight configuration (\texttt{dmodel}=128, \texttt{nhead}=4) that balances accuracy, efficiency, and employability. These results confirm the potential of Transformer-based acceleration for real-time optimal control. Future work includes extending Quattro to more complex robotic systems, investigating adaptive prompt tuning, and exploring integration with learning-based control for greater robustness in dynamic environments.




\bibliographystyle{IEEEtran} 
\bibliography{references} 

\begin{thebibliography}{10}
\providecommand{\url}[1]{#1}
\csname url@samestyle\endcsname
\providecommand{\newblock}{\relax}
\providecommand{\bibinfo}[2]{#2}
\providecommand{\BIBentrySTDinterwordspacing}{\spaceskip=0pt\relax}
\providecommand{\BIBentryALTinterwordstretchfactor}{4}
\providecommand{\BIBentryALTinterwordspacing}{\spaceskip=\fontdimen2\font plus
\BIBentryALTinterwordstretchfactor\fontdimen3\font minus \fontdimen4\font\relax}
\providecommand{\BIBforeignlanguage}[2]{{%
\expandafter\ifx\csname l@#1\endcsname\relax
\typeout{** WARNING: IEEEtran.bst: No hyphenation pattern has been}%
\typeout{** loaded for the language `#1'. Using the pattern for}%
\typeout{** the default language instead.}%
\else
\language=\csname l@#1\endcsname
\fi
#2}}
\providecommand{\BIBdecl}{\relax}
\BIBdecl

\bibitem{bledt2018cheetah}
G.~Bledt, M.~J. Powell, B.~Katz, J.~Di~Carlo, P.~M. Wensing, and S.~Kim, ``Mit cheetah 3: Design and control of a robust, dynamic quadruped robot,'' in \emph{2018 IEEE/RSJ International Conference on Intelligent Robots and Systems (IROS)}.\hskip 1em plus 0.5em minus 0.4em\relax IEEE, 2018, pp. 2245--2252.

\bibitem{jeon2022online}
S.~H. Jeon, S.~Kim, and D.~Kim, ``Online optimal landing control of the mit mini cheetah,'' in \emph{2022 International Conference on Robotics and Automation (ICRA)}.\hskip 1em plus 0.5em minus 0.4em\relax IEEE, 2022, pp. 178--184.

\bibitem{romualdi2022online}
G.~Romualdi, S.~Dafarra, G.~L'Erario, I.~Sorrentino, S.~Traversaro, and D.~Pucci, ``Online non-linear centroidal mpc for humanoid robot locomotion with step adjustment,'' in \emph{2022 International Conference on Robotics and Automation (ICRA)}.\hskip 1em plus 0.5em minus 0.4em\relax IEEE, 2022, pp. 10\,412--10\,419.

\bibitem{lindqvist2020nonlinear}
B.~Lindqvist, S.~S. Mansouri, A.-a. Agha-mohammadi, and G.~Nikolakopoulos, ``Nonlinear mpc for collision avoidance and control of uavs with dynamic obstacles,'' \emph{IEEE robotics and automation letters}, vol.~5, no.~4, pp. 6001--6008, 2020.

\bibitem{nguyen2023mpc}
D.~G. Nguyen, S.~Park, J.~Park, D.~Kim, J.~S. Eo, and K.~Han, ``An mpc approximation approach for adaptive cruise control with reduced computational complexity and low memory footprint,'' \emph{IEEE Transactions on Intelligent Vehicles}, vol.~9, no.~2, pp. 3154--3167, 2023.

\bibitem{tassa2012synthesis}
Y.~Tassa, T.~Erez, and E.~Todorov, ``Synthesis and stabilization of complex behaviors through online trajectory optimization,'' in \emph{2012 IEEE/RSJ International Conference on Intelligent Robots and Systems}.\hskip 1em plus 0.5em minus 0.4em\relax IEEE, 2012, pp. 4906--4913.

\bibitem{tassa2014control}
Y.~Tassa, N.~Mansard, and E.~Todorov, ``Control-limited differential dynamic programming,'' in \emph{2014 IEEE International Conference on Robotics and Automation (ICRA)}.\hskip 1em plus 0.5em minus 0.4em\relax IEEE, 2014, pp. 1168--1175.

\bibitem{neunert2017trajectory}
M.~Neunert, F.~Farshidian, A.~W. Winkler, and J.~Buchli, ``Trajectory optimization through contacts and automatic gait discovery for quadrupeds,'' \emph{IEEE Robotics and Automation Letters}, vol.~2, no.~3, pp. 1502--1509, 2017.

\bibitem{neunert2018whole}
M.~Neunert, M.~St{\"a}uble, M.~Giftthaler, C.~D. Bellicoso, J.~Carius, C.~Gehring, M.~Hutter, and J.~Buchli, ``Whole-body nonlinear model predictive control through contacts for quadrupeds,'' \emph{IEEE Robotics and Automation Letters}, vol.~3, no.~3, pp. 1458--1465, 2018.

\bibitem{howell2019altro}
T.~A. Howell, B.~E. Jackson, and Z.~Manchester, ``Altro: A fast solver for constrained trajectory optimization,'' in \emph{2019 IEEE/RSJ International Conference on Intelligent Robots and Systems (IROS)}.\hskip 1em plus 0.5em minus 0.4em\relax IEEE, 2019, pp. 7674--7679.

\bibitem{zhu2024convergent}
J.~Zhu, J.~J. Payne, and A.~M. Johnson, ``Convergent ilqr for safe trajectory planning and control of legged robots,'' in \emph{2024 IEEE International Conference on Robotics and Automation (ICRA)}.\hskip 1em plus 0.5em minus 0.4em\relax IEEE, 2024, pp. 8051--8057.

\bibitem{plancher2020performance}
B.~Plancher and S.~Kuindersma, ``A performance analysis of parallel differential dynamic programming on a gpu,'' in \emph{Algorithmic Foundations of Robotics XIII: Proceedings of the 13th Workshop on the Algorithmic Foundations of Robotics 13}.\hskip 1em plus 0.5em minus 0.4em\relax Springer, 2020, pp. 656--672.

\bibitem{chen2020survey}
Y.~Chen, Y.~Xie, L.~Song, F.~Chen, and T.~Tang, ``A survey of accelerator architectures for deep neural networks,'' \emph{Engineering}, vol.~6, no.~3, pp. 264--274, 2020.

\bibitem{cai2024medusa}
T.~Cai, Y.~Li, Z.~Geng, H.~Peng, J.~D. Lee, D.~Chen, and T.~Dao, ``Medusa: Simple llm inference acceleration framework with multiple decoding heads,'' \emph{arXiv preprint arXiv:2401.10774}, 2024.

\bibitem{pellegrini2020multiple}
E.~Pellegrini and R.~P. Russell, ``A multiple-shooting differential dynamic programming algorithm. part 1: Theory,'' \emph{Acta Astronautica}, vol. 170, pp. 686--700, 2020.

\bibitem{giftthaler2018family}
M.~Giftthaler, M.~Neunert, M.~St{\"a}uble, J.~Buchli, and M.~Diehl, ``A family of iterative gauss-newton shooting methods for nonlinear optimal control,'' in \emph{2018 IEEE/RSJ International Conference on Intelligent Robots and Systems (IROS)}.\hskip 1em plus 0.5em minus 0.4em\relax IEEE, 2018, pp. 1--9.

\bibitem{lee2022gpu}
Y.~Lee, M.~Cho, and K.-S. Kim, ``Gpu-parallelized iterative lqr with input constraints for fast collision avoidance of autonomous vehicles,'' in \emph{2022 IEEE/RSJ International Conference on Intelligent Robots and Systems (IROS)}.\hskip 1em plus 0.5em minus 0.4em\relax IEEE, 2022, pp. 4797--4804.

\bibitem{dai2024parallel}
C.~Dai, J.~Su, and J.~Wang, ``A parallel iterative linear quadratic controller for autonomous trajectory optimization,'' in \emph{2024 43rd Chinese Control Conference (CCC)}.\hskip 1em plus 0.5em minus 0.4em\relax IEEE, 2024, pp. 4573--4578.

\bibitem{li2023unified}
H.~Li, W.~Yu, T.~Zhang, and P.~M. Wensing, ``A unified perspective on multiple shooting in differential dynamic programming,'' in \emph{2023 IEEE/RSJ International Conference on Intelligent Robots and Systems (IROS)}.\hskip 1em plus 0.5em minus 0.4em\relax IEEE, 2023, pp. 9978--9985.

\bibitem{hong2023physics}
S.~H. Hong, J.~Ou, and Y.~Wang, ``Physics-guided neural network and gpu-accelerated nonlinear model predictive control for quadcopter,'' \emph{Neural Computing and Applications}, vol.~35, no.~1, pp. 393--413, 2023.

\bibitem{celestini2024transformer}
D.~Celestini, D.~Gammelli, T.~Guffanti, S.~D'Amico, E.~Capello, and M.~Pavone, ``Transformer-based model predictive control: Trajectory optimization via sequence modeling,'' \emph{IEEE Robotics and Automation Letters}, 2024.

\bibitem{zinage2024transformermpc}
V.~Zinage, A.~Khalil, and E.~Bakolas, ``Transformermpc: Accelerating model predictive control via transformers,'' \emph{arXiv preprint arXiv:2409.09266}, 2024.

\bibitem{elman1990finding}
J.~L. Elman, ``Finding structure in time,'' \emph{Cognitive science}, vol.~14, no.~2, pp. 179--211, 1990.

\bibitem{hochreiter1997long}
S.~Hochreiter and J.~Schmidhuber, ``Long short-term memory,'' \emph{Neural computation}, vol.~9, no.~8, pp. 1735--1780, 1997.

\bibitem{lipton2015critical}
Z.~C. Lipton, J.~Berkowitz, and C.~Elkan, ``A critical review of recurrent neural networks for sequence learning,'' \emph{arXiv preprint arXiv:1506.00019}, 2015.

\bibitem{vaswani2017attention}
A.~Vaswani, N.~Shazeer, N.~Parmar, J.~Uszkoreit, L.~Jones, A.~N. Gomez, {\L}.~Kaiser, and I.~Polosukhin, ``Attention is all you need,'' \emph{Advances in neural information processing systems}, vol.~30, 2017.

\bibitem{radford2018improving}
A.~Radford, K.~Narasimhan, T.~Salimans, I.~Sutskever \emph{et~al.}, ``Improving language understanding by generative pre-training,'' 2018.

\bibitem{devlin2019bert}
J.~Devlin, M.-W. Chang, K.~Lee, and K.~Toutanova, ``Bert: Pre-training of deep bidirectional transformers for language understanding,'' in \emph{Proceedings of the 2019 conference of the North American chapter of the association for computational linguistics: human language technologies, volume 1 (long and short papers)}, 2019, pp. 4171--4186.

\bibitem{soydaner2022attention}
D.~Soydaner, ``Attention mechanism in neural networks: where it comes and where it goes,'' \emph{Neural Computing and Applications}, vol.~34, no.~16, pp. 13\,371--13\,385, 2022.

\bibitem{nvidia_dlss4}
\BIBentryALTinterwordspacing
{NVIDIA Corporation}. (2025) {DLSS 4 Introduces Multi Frame Generation and Transformer-Based AI Models}. Accessed: 2025-03-19. [Online]. Available: \url{https://www.nvidia.com/en-us/geforce/news/dlss4-multi-frame-generation-ai-innovations/}
\BIBentrySTDinterwordspacing

\bibitem{jouppi2017datacenter}
N.~P. Jouppi, C.~Young, N.~Patil, D.~Patterson, G.~Agrawal, R.~Bajwa, S.~Bates, S.~Bhatia, N.~Boden, A.~Borchers \emph{et~al.}, ``In-datacenter performance analysis of a tensor processing unit,'' in \emph{Proceedings of the 44th annual international symposium on computer architecture}, 2017, pp. 1--12.

\bibitem{li2020ftrans}
B.~Li, S.~Pandey, H.~Fang, Y.~Lyv, J.~Li, J.~Chen, M.~Xie, L.~Wan, H.~Liu, and C.~Ding, ``Ftrans: energy-efficient acceleration of transformers using fpga,'' in \emph{Proceedings of the ACM/IEEE International Symposium on Low Power Electronics and Design}, 2020, pp. 175--180.

\bibitem{zhao2022fpga}
Z.~Zhao, R.~Cao, K.-F. Un, W.-H. Yu, P.-I. Mak, and R.~P. Martins, ``An fpga-based transformer accelerator using output block stationary dataflow for object recognition applications,'' \emph{IEEE Transactions on Circuits and Systems II: Express Briefs}, vol.~70, no.~1, pp. 281--285, 2022.

\bibitem{tunyasuvunakool2020}
S.~Tunyasuvunakool, A.~Muldal, Y.~Doron, S.~Liu, S.~Bohez, J.~Merel, T.~Erez, T.~Lillicrap, N.~Heess, and Y.~Tassa, ``dm\_control: Software and tasks for continuous control,'' \emph{Software Impacts}, vol.~6, p. 100022, 2020.

\bibitem{menagerie2022github}
K.~Zakka, Y.~Tassa, and {MuJoCo Menagerie Contributors}, ``Mujoco menagerie: A collection of high-quality simulation models for mujoco,'' \url{http://github.com/google-deepmind/mujoco_menagerie}, 2022, accessed: 2025-03-31.

\bibitem{chen2024allo}
\BIBentryALTinterwordspacing
H.~Chen, N.~Zhang, S.~Xiang, Z.~Zeng, M.~Dai, and Z.~Zhang, ``Allo: A programming model for composable accelerator design,'' \emph{Proc. ACM Program. Lang.}, vol.~8, no. PLDI, jun 2024. [Online]. Available: \url{https://doi.org/10.1145/3656401}
\BIBentrySTDinterwordspacing

\bibitem{dong2018speech}
L.~Dong, S.~Xu, and B.~Xu, ``Speech-transformer: a no-recurrence sequence-to-sequence model for speech recognition,'' in \emph{2018 IEEE international conference on acoustics, speech and signal processing (ICASSP)}.\hskip 1em plus 0.5em minus 0.4em\relax IEEE, 2018, pp. 5884--5888.

\bibitem{fuad2023survey}
K.~A.~A. Fuad and L.~Chen, ``A survey on sparsity exploration in transformer-based accelerators,'' \emph{Electronics}, vol.~12, no.~10, p. 2299, 2023.

\bibitem{liu2022dynamic}
L.~Liu, Z.~Qu, Z.~Chen, F.~Tu, Y.~Ding, and Y.~Xie, ``Dynamic sparse attention for scalable transformer acceleration,'' \emph{IEEE Transactions on Computers}, vol.~71, no.~12, pp. 3165--3178, 2022.

\bibitem{park2020optimus}
J.~Park, H.~Yoon, D.~Ahn, J.~Choi, and J.-J. Kim, ``Optimus: Optimized matrix multiplication structure for transformer neural network accelerator,'' \emph{Proceedings of Machine Learning and Systems}, vol.~2, pp. 363--378, 2020.

\bibitem{yang2022efa}
X.~Yang and T.~Su, ``Efa-trans: An efficient and flexible acceleration architecture for transformers,'' \emph{Electronics}, vol.~11, no.~21, p. 3550, 2022.

\bibitem{todorov2012mujoco}
E.~Todorov, T.~Erez, and Y.~Tassa, ``Mujoco: A physics engine for model-based control,'' in \emph{2012 IEEE/RSJ International Conference on Intelligent Robots and Systems}.\hskip 1em plus 0.5em minus 0.4em\relax IEEE, 2012, pp. 5026--5033.

\bibitem{mckay2000comparison}
M.~D. McKay, R.~J. Beckman, and W.~J. Conover, ``A comparison of three methods for selecting values of input variables in the analysis of output from a computer code,'' \emph{Technometrics}, vol.~42, no.~1, pp. 55--61, 2000.

\bibitem{cvxpy}
\BIBentryALTinterwordspacing
S.~Diamond and S.~Boyd, ``{CVXPY}: A {P}ython-embedded modeling language for convex optimization,'' \emph{Journal of Machine Learning Research}, 2016, to appear. [Online]. Available: \url{https://stanford.edu/~boyd/papers/pdf/cvxpy_paper.pdf}
\BIBentrySTDinterwordspacing

\end{thebibliography}

\end{document}